\documentclass[twocolumn,10pt,aps,pra,showpacs,superscriptaddress,nobalancelastpage,longbibliography,nofootinbib,floatfix]{revtex4-2}

\usepackage{graphicx}
\usepackage{dcolumn}
\usepackage{bm}

\usepackage{braket} 
\usepackage{graphicx} 
\usepackage{amssymb}
\usepackage{amsmath}
\DeclareMathOperator*{\argmax}{arg\,max}
\DeclareMathOperator*{\argmin}{arg\,min}
\usepackage{algorithm}
\usepackage[noend]{algpseudocode}
\usepackage{graphicx}
\usepackage{subcaption}
\usepackage{wrapfig}
\usepackage{adjustbox}
\usepackage{relsize}
\usepackage{hyperref}
\usepackage{xcolor}
\usepackage{svg}
\usepackage{natbib}

\usepackage{tikz} 
\usetikzlibrary{quantikz} 

\usepackage{hhline}

\begin{document}

\title{Radio Signal Classification by Adversarially Robust Quantum Machine Learning}

\author{Yanqiu Wu}
\email{autumn.wu@data61.csiro.au}
\affiliation{Data61, CSIRO, Marsfield, NSW, 2122, Australia}

\author{Eromanga Adermann}
\affiliation{Data61, CSIRO, Marsfield, NSW, 2122, Australia}

\author{Chandra Thapa}
\affiliation{Data61, CSIRO, Marsfield, NSW, 2122, Australia}

\author{Seyit Camtepe}
\affiliation{Data61, CSIRO, Marsfield, NSW, 2122, Australia}

\author{Hajime Suzuki}
\affiliation{Data61, CSIRO, Marsfield, NSW, 2122, Australia}

\author{Muhammad Usman}
\affiliation{Data61, CSIRO, Clayton, VIC, 3168, Australia}
\affiliation{School of Physics, The University of Melbourne, Parkville, 3010, Victoria Australia}

\begin{abstract}
    Radio signal classification plays a pivotal role in identifying the modulation scheme used in received radio signals, which is essential for demodulation and proper interpretation of the transmitted information. Researchers have underscored the high susceptibility of ML algorithms for radio signal classification to adversarial attacks. Such vulnerability could result in severe consequences, including misinterpretation of critical messages, interception of classified information, or disruption of communication channels. Recent advancements in quantum computing have revolutionized theories and implementations of computation, bringing the unprecedented development of Quantum Machine Learning (QML). It is shown that quantum variational classifiers (QVCs) provide notably enhanced robustness against classical adversarial attacks in image classification. However, no research has yet explored whether QML can similarly mitigate adversarial threats in the context of radio signal classification. This work applies QVCs to radio signal classification and studies their robustness to various adversarial attacks. We also propose the novel application of the approximate amplitude encoding (AAE) technique to encode radio signal data efficiently. Our extensive simulation results present that attacks generated on QVCs transfer well to CNN models, indicating that these adversarial examples can fool neural networks that they are not explicitly designed to attack. However, the converse is not true. QVCs primarily resist the attacks generated on CNNs. Overall, with comprehensive simulations, our results shed new light on the growing field of QML by bridging knowledge gaps in QAML in radio signal classification and uncovering the advantages of applying QML methods in practical applications. 
\end{abstract}

\maketitle

\section{Introduction}
Radio frequency allocation for mobile and Wi-Fi services has expanded significantly from the 1990s to the 2020s, increasing bandwidth from tens of megahertz to multiple gigahertz. This growing demand for wireless service frequency faces the challenge of radio frequency being a finite resource, necessitating efficient usage technologies \cite{OUGHTON2021102127}.
Modern digital radio services utilize diverse modulation and coding schemes (MCS) to optimize data transfer rates. These schemes require extra communication between the transmitter and receiver to inform receivers of the forthcoming modulation type ahead of every frame, consuming additional bandwidth. Minimizing this overhead is crucial, particularly in bandwidth-limited situations.

Automatic modulation classification (AMC)~\cite{9576081} solves the above-mentioned problem where the receiver classifies the modulation scheme of the received radio packet without requiring control overhead solely based on the signal's physical features. This process is crucial in wireless communication systems applications such as signal intelligence, surveillance, cognitive radio, and dynamic spectrum access, and it is essentially a classification problem.
While various AMC methods have been proposed in the past, machine learning (ML)-based methods have been shown to excel in recent years \cite{9576081}. However, despite their ability to obtain high classification accuracy, researchers have underscored the increased susceptibility of ML algorithms for radio signal classification to adversarial attacks. For example, Sadeghi and Larsson \cite{Sadeghi2018AdversarialAO} used a variant of fast gradient sign method (FGSM) \cite{iclr-fgsm} to perform an adversarial ML attack on CNN-based modulation classification and successfully showed a considerable drop in classification accuracy. More recently, Usama et al. \cite{8881843} evaluated Carlini \& Wagner (C-W) attack \cite{carlini2017evaluating} on nine ML-based signal classifiers to highlight the vulnerability of these classifiers to adversarial perturbations. Such vulnerability of these signal models to adversarial attacks could result in severe consequences, including misinterpretation of critical messages, interception of classified information, or disruption of communication channels in wireless communications.

Different defense strategies have been proposed in response to the increasingly sophisticated and strong attack methods, such as adversarial training, model ensembles, and GAN \citep{goodfellow2014generative}. While these studies of adversarial machine learning (AML) remain open, there has been growing interest recently in how quantum machine learning (QML) will fare against adversarial attacks \citep{PhysRevResearch.2.033212, PhysRevResearch.5.023186, West_2023}. The advancements in quantum computing over the last few years have revolutionized theories and implementations of computation, which brings the intersection of quantum computing and machine learning to the unprecedented development of the new field of QML\citep{Biamonte_2017, Ciliberto_2018}. Research has been 
conducted to show striking opportunities to enhance, speed up, and innovate machine learning with quantum methods \citep{Dunjko_2018}. This field proliferates with notable progress and brings attention to the research area of quantum machine learning under different adversarial scenarios, which is called quantum adversarial machine learning (QAML) \cite{West_2023}. For example, West et al. \cite{PhysRevResearch.5.023186} presented a theoretical study that adversarial examples generated by quantum attacks target robust features that allow the QAML algorithm to generate strong attacks for high-performance 
classical models. Meanwhile, it is shown that QVCs provide a notably enhanced robustness against classical adversarial attacks in image classification by learning features not detected by the classical neural networks \cite{West_2023}. Despite recent notable progress, the field of QAML is still in its infancy. Many important issues remain largely unexplored, and most research works mainly focus on computer vision for image classification \citep{PhysRevResearch.2.033212, zhang2019adversarial, PhysRevResearch.5.023186, West_2023}. It is also worth noting that while QML has demonstrated improved robustness against adversarial attacks in image classification, no research has yet explored whether QML can similarly mitigate adversarial threats in the context of radio signal classification.

\begin{figure*}
    \centering
    \includegraphics[width=0.99\linewidth]{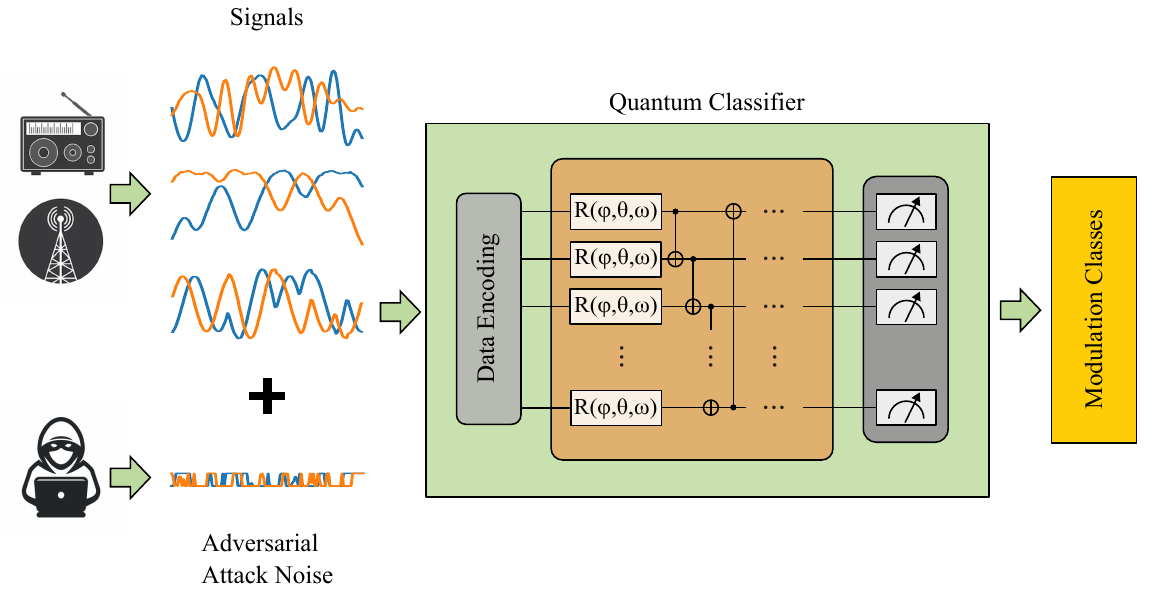}
    \caption{An illustration of Quantum Adversarial Machine Learning and our quantum classifier architecture. We implement quantum variational circuit (QVC) as the quantum classifier. The variational component of the circuit is made up of layers of 3 rotation gates per wire, followed by a Controlled-NOT gate connecting two adjacent wires. The rotation angles of the gates constitute the trainable parameters of the QVC. To simulate an adversarial attack, attackers generate adversarial noise which is then blended with the original radio signal data.}
    \label{fig:architect}
\end{figure*}

\paragraph*{\textbf{Our contributions:}}
This paper develops and implements a quantum machine learning algorithm to radio signal classification and studies its robustness to various adversarial attacks. Figure \ref{fig:architect} schematically illustrates our framework to investigate QAML for radio signals. In this regard, we compare the performance of quantum variational classifiers (QVCs) and widely used classical convolutional neural networks (CNNs) for radio signal classification. \textit{To the best of our knowledge, this is the first work to study QVCs under adversarial attacks for radio signal classification.} We carried out an extensive set of quantum and classical simulations across concrete radio-frequency examples and synthetic Fourier series waveform dataset, which covers different scenarios with different state-of-the-art attack strategies, including FGSM \citep{iclr-fgsm}, projected gradient descent (PGD) \citep{DBLP:conf/iclr/MadryMSTV18}, and universal adversarial perturbations (UAP) \citep{Sadeghi2018AdversarialAO} in the white-box and black-box attack setting. In white-box attacks, the adversary has the full knowledge of the classifier, while in black-box attacks, the adversary does not know the classifier.
Additionally, we explore the following research questions (RQs), delving deeper into adversarial aspects:
\begin{enumerate}
    \item [RQ1] \textit{Are adversarial examples transferable between quantum classifier and classical classifier?}\\
    Our results highlight that adversarial examples generated by attacking a classical model fail to fool quantum classifiers. However, perturbations generated by quantum attacks transfer well to classical models and are capable of deceiving classical models. 

    \item [RQ2] \textit{What is the attack stealthiness of the adversarial samples generated by quantum classifiers and classical classifiers?}\\
    Under different attack levels for the FGSM and PGD attacks, the adversarial examples generated by QVCs are generally more imperceptible than those generated by CNN. For UAP attacks, adversarial examples generated by CNN are considered slightly more imperceptible than those generated by QVC.

    \item [RQ3] \textit{Can we improve encoding to improve the robustness of quantum classifiers?}\\
    We propose the novel application of the approximate amplitude encoding (AAE) technique for encoding radio signal data. With 5 layers of AAE layers, we significantly reduce the number of gates required for encoding from 973 to 155 by sacrificing 3.7\% accuracy. With AAE, the QVC model is still very robust against black-box adversarial attacks. The reduction of quantum resources will greatly help in reducing the detrimental impact of noise on QML fidelity and ultimately on fault-tolerant implementation.

    \item [RQ4] \textit{How the quantum noise would affect the classification tasks?}\\
    The introduction of depolarisation noise resulted in a significant decrease in classification accuracy. However, it did not impact the QVC model's robustness against black-box adversarial attacks. 
    
\end{enumerate}

\section{Related Works}
\label{sec:relatedworks}
Data-driven machine learning has achieved remarkable success in a wide range of applications. During the process, people become more concerned with the security of the applications, especially in security-critical tasks such as self-driving cars, medical diagnostics and autonomous weapons. \citep{miller2019adversarial, Biggio_2018}. To address these concerns, an emergent field of adversarial machine learning has been motivated to research the vulnerability of a variety of machine learning algorithms under different adversarial attack settings and to develop defense strategies to fare against adversarial attacks \citep{10.5555/3350831, 7478523}. With the advancement in quantum computing and the increasing interest in quantum machine learning, researchers have taken further steps to study the robustness of quantum classifiers and possible defense strategies. Lu et al. \cite{PhysRevResearch.2.033212} carried out typical adversarial attacks\citep{iclr-fgsm,kurakin2017adversarial2,DBLP:conf/iclr/MadryMSTV18,Chen_2017} in the classical setting on a standard quantum variational circuits (QVCs) for both classical MNIST\citep{726791} and quantum datasets \citep{jiang2021adversarial}. The results show that under different white-box attack scenarios, QVCs can be fooled with adversarially attacked images that are very similar to the original clean images, even in the relatively simple case of binary classification. In addition to the simulations, Ren et al. \cite{Ren_2022} carried out adversarial studies of QVCs on quantum hardware which is fundamentally impacted by noise.
The results found that despite the virtual similarities between adversarial examples and their corresponding clean images, QVCs trained on real quantum hardware can still be fooled, demonstrating the real negative effects of adversarial attacks on current noisy hardware. However, incorporating an adversarial training process significantly increases their resilience to these perturbations.
More recently, West et al. \cite{PhysRevResearch.5.023186} benchmarked the quantum ML networks for a range of image datasets \cite{726791, DBLP:journals/corr/abs-1708-07747, yang2015facial} and study its robustness through a variety of adversarial attacks. Through systematic simulations, researchers found that QVCs display remarkable resilience to the adversarial attacks generated on classical neural networks for image classification in a black-box setting. In contrast, adversarial examples constructed by carrying out white-box attacks on the QVCs tend to transfer well to the classical models. Despite these QAML studies on image classification tasks, it is still an open question how QML and QAML will perform when applied and tested on signal datasets.

\section{Adversarial Attacks for Wireless Communication Systems}
\label{sec:adversarialattack}
This section provides a detailed procedure for performing different adversarial attacks on modulation classification. We consider a wireless communication system where we have a transmitter, a receiver, and an attacker. 

In automatic modulation classification, the transmitter transmits signals with one specific modulation type. The receiver receives the wireless signal and applies a pre-trained classifier to classify the modulation type used at the transmitter. The attacker launches an attack by transmitting a signal to create a low-power perturbation and cause misclassification at the receiver. 

We denote the modulation classifier at the receiver by $f(.;\bm{\theta}): \chi \to \mathbb{R}^{C} $, where $\bm{\theta}$ is the set of model parameters of the classifier, $\chi \in \mathbb{R}^{p} $ is the input domain where $p$ is the dimension of the inputs and $C$ is the number of modulation types. For every input $\mathbf{x}$, the classifier assigns a modulation type $\hat{l}(\mathbf{x},\bm{\theta})= \argmax_{k}f_{k}(\mathbf{x},\bm{\theta})$ where $f_k(\mathbf{x},\bm{\theta})$ is the output of the classifier corresponding to the $k$th class. If the adversary transmits a perturbation $r_{\mathbf{x}}$ which causes misclassification at the receiver, we have $\hat{l}(\mathbf{x},\bm{\theta}) \neq \hat{l}(\mathbf{x}+r_{\mathbf{x}},\bm{\theta})$ and $\mathbf{x}+r_{\mathbf{x}} \in \chi$. Given these definitions, the adversary obtains the adversarial perturbation $r_{\mathbf{x}}$ for the input $\mathbf{x}$ and classifier $f$ by solving the following optimization problem \citep{Sadeghi2018AdversarialAO}:
\begin{equation}
\label{eq1}
    \argmin_{r_{\mathbf{x}}}  \quad  \lVert r_{\mathbf{x}} \rVert_2 \quad s.t. \quad \hat{l}(\mathbf{x},\bm{\theta}) \neq \hat{l}(\mathbf{x}+r_{\mathbf{x}},\bm{\theta})
\end{equation}
In practice, solving (\ref{eq1}) is difficult. Hence, different methods, such as fast gradient methods (FGM), have been proposed to approximate the adversarial perturbation. There are also two variants of FGM: targeted FGM and non-targeted FGM \citep{Sadeghi2018AdversarialAO, 8881843}. In a targeted FGM attack, the adversary tries to generate a perturbation that causes the classifier at the receiver to have a specific misclassification, e.g., classifying QPSK modulation as 8PSK modulation. In a non-targeted FGM attack, the adversary generates a perturbation that causes any misclassification independent of the target label. In this paper, we focus on non-targeted adversarial attack algorithms. 

In addition to the targeted and non-targeted categories, adversarial attacks can also be categorized along other dimensions, such as white-box and black-box attacks \citep{iclr-fgsm, DBLP:journals/corr/SzegedyZSBEGF13}. The attack category is determined based on the amount of knowledge that the adversary has about the model. In a white-box attack, the adversary is considered to have the full knowledge of the classifier. In contrast, in a black-box attack, the adversary has no or limited knowledge of the classifier.  

In this paper, we employ the novel application of QVCs on radio signal classification and study its robustness to various adversarial attacks. For the white-box adversarial attacks, the adversary is assumed to know the classifier's model architecture and input data at the receiver. We will further explain the different types of attacks we experiment with in Section \ref{subsec:whitebox} and \ref{subsec:blackbox}.

\subsection{White-box attack}
\label{subsec:whitebox}
\subsubsection{Fast Gradient Sign Method}
\label{sub:fgsm}

FGM is a commonly used approach for solving (\ref{eq1}) \citep{iclr-fgsm}.  Let $J(\bm{\theta}, \mathbf{x}, \mathbf{y})$ be the cost function used to train the modulation classifier $f(.;\bm{\theta})$, where $\mathbf{y}$ is the label vector. Goodfellow et al. \cite{iclr-fgsm} proposed fast gradient sign method(FGSM), which utilized $L_{\infty}$ bounded constraint and the sign function to get the gradient direction. It then takes a small step in the gradient direction that will maximize the loss until suitable perturbation is reached:
\begin{equation}
    \eta = \epsilon\times \text{sign}(\nabla_{\mathbf{x}} J(\bm{\theta}, \mathbf{x}, \mathbf{y}))
\end{equation}
FGSM is proposed and widely applied in Computer Vision. It is shown that FGSM effectively causes various classifiers to misclassify their input \citep{iclr-fgsm}. In this paper, we adopted this method and applied it to radio signal data \citep{grcon} and studied how quantum variational circuits and classical CNN model react to the FGSM attack. 

\subsubsection{Projected Gradient Descent}
As mentioned in the previous section, FGSM is a simple one-step scheme to generate powerful adversarial attacks. 
Madry et al. \cite{DBLP:conf/iclr/MadryMSTV18} proposed an improvement of the multi-step variant to FGSM, which is referred to as projected gradient descent (PGD). PGD first uniforms the random perturbation as the initialization and then generates the perturbation by running several iterations of FGSM. Formally, the iterative process is presented as follows:
\begin{equation}
 \mathbf{x}^{t+1} = \Pi_{\mathbf{x},\epsilon} \{\mathbf{x}^t + \alpha \text{sign} \nabla_{\mathbf{x}} J(\bm{\theta}, \mathbf{x}, \mathbf{y}) \}, \quad \mathbf{x}^0 = \mathbf{x}
\end{equation} 
where $\Pi$ denotes the projection operator, which clips the adversarial examples around the predefined perturbation range. $\alpha$ represents the gradient step size.

\subsubsection{Universal Adversarial Perturbations}
FGSM attacks and PGD attacks have some limitations. First of all, the adversary needs to know the full 
input data. Secondly, each input data $\mathbf{x}$ is perturbed by its corresponding noise $r_{\mathbf{x}}$. In other words, the adversary is always required to know the classifier's input, which is impractical. Sadeghi and Larsson \cite{Sadeghi2018AdversarialAO} address these limitations by introducing an input-agnostic algorithm. Specifically, instead of generating perturbations $r_{\mathbf{x}}$ corresponding to each input $\mathbf{x}$, a universal adversarial perturbation $r$ is generated to fool the classifier with high probability. We refer to the algorithm as Universal Adversarial Perturbations (UAP).

The UAP algorithm \citep{Sadeghi2018AdversarialAO} utilizes principal component analysis (PCA) to craft the perturbations. Assume the adversary collects an arbitrary subset of inputs $\{\mathbf{x}_1, \dots, \mathbf{x}_N\}$, and the corresponding perturbation directions $\{n_{\mathbf{x}_1}, \dots, n_{\mathbf{x}_N}\}$ is known, where $n_{\mathbf{x}_i} = \nabla_{\mathbf{x}_i} L(\bm{\theta}, \mathbf{x}_i, \mathbf{y}^\text{true})/ \lVert  \nabla_{\mathbf{x}_i} L(\bm{\theta}, \mathbf{x}_i, \mathbf{y}^\text{true}) \rVert_2$ \citep{Sadeghi2018AdversarialAO}. We stack $n_{\mathbf{x}_1}$ to $n_{\mathbf{x}_N}$ into a matrix, then perform PCA to find the first principal component of the matrix with the largest eigenvalue. The direction of the first principal component will account for the most variability in $\{ n_{\mathbf{x}_1}, \dots, n_{\mathbf{x}_N}\}$ which is proposed to be used as the direction of the UAP. A pseudo-code is provided in Algorithm \ref{alg:uap}.

\begin{algorithm}[H]
\caption{PCA-based UAP algorithm}
\label{alg:uap}
\hspace*{\algorithmicindent} \textbf{Input:} a random subset of input data $\{\mathbf{x}_1,\dots,\mathbf{x}_N\}$ and their corresponding labels, the classifier $f(.,\bm{\theta})$ and the maximum allowed perturbation norm $p_\text{max}$
\begin{algorithmic}[1]
\State Evaluate the matrix $X^{N \times p} = [n_{\mathbf{x}_1}, \dots, n_{\mathbf{x}_N}]^T$
\State Compute the first principal component of $X$ and denote it by $v_1$, i.e., $X=U \Sigma V^T$ and $v_1 = V e_1$.
\State $r = p_\text{max} v_1$
\State return $r$
\end{algorithmic}
\end{algorithm}

\subsection{Black-box Attack}
\label{subsec:blackbox}
In a black-box attack scenario, we relax the information about the classifier at the receiver. We assume no information about the internal structures of the classifiers and the learning algorithms. To achieve this assumption, we utilize the transferability property of adversarial examples \citep{szegedy2014intriguing, iclr-fgsm}. This property states that, with high probability, adversarial example crafted to fool one specific learning model can fool another model even if their architecture differ greatly or they are trained on different sets of training data \citep{DBLP:conf/iclr/MadryMSTV18}.  

In this paper, we use the transferability of adversarial examples in a more exotic setting for black-box attacks. That is, to craft adversarial examples for the classical CNN model, we first use the quantum classifier QVC as a substitute model, derive adversarial examples from it the same way as in a white-box attack, and then apply them on the CNN model to investigate whether they transfer to the classical classifier. Similarly, to craft perturbations for quantum classifier, we use the CNN model as a substitute model and derive adversarial examples from it the same way as in white-box attack and then apply them on the QVC and QVC with Approximate Amplitude Encoding for investigation.

\begin{figure*}
    \includegraphics[width=\textwidth]{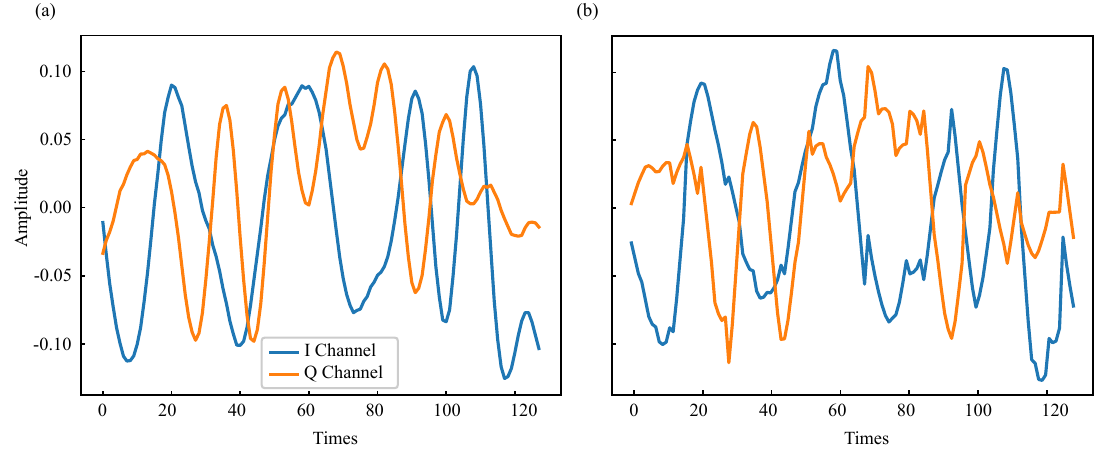}
    \caption{Visualization of signal data. (a) shows the visualization of an example signal data when encoded with exact amplitude encoding. (b) shows the visualization of the same signal data when encoded with the approximate amplitude encoding method. }
    \label{fig:approx-encode}
\end{figure*}

\begin{figure*}
    \centering
    \includegraphics[width=\textwidth]{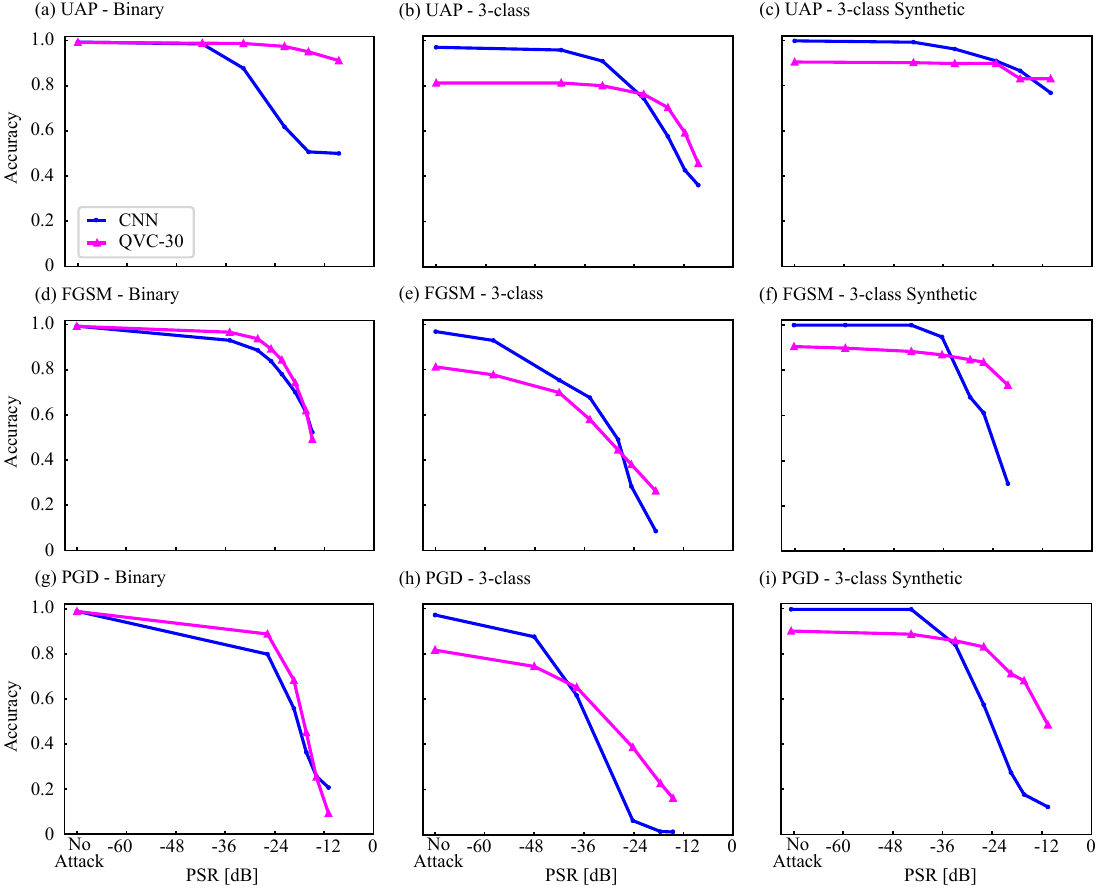}
    \caption{The performance achieved by QVC (purple) and CNN (blue) in the case of white-box UAP, FGSM, and PGD attacks on the GNU radio ML dataset and our synthetic dataset as a function of attack strength defined as perturbation-to-signal ratio (PSR). Figures (a) to (c) represent the performance under white-box UAP attack for binary RML dataset, 3-class RML dataset, and synthetic dataset, respectively. Figures (d) to (f) represent the performance under white-box FGSM attack for binary RML dataset, 3-class RML dataset, and synthetic dataset, respectively. Figures (g) to (i) represent the performance under white-box PGD attack for binary RML dataset, 3-class RML dataset, and synthetic dataset, respectively. For each sub-figure, the left most dots represent the initial accuracy of the models without any attacks. Across different attacks and datasets, we can see that even though CNN usually has the highest initial accuracy, QVC is generally more robust. However, QVC becomes vulnerable as the attack strength increases.}
    \label{fig:wb-results}
\end{figure*}

\begin{figure*}[htbp]
    \centering
    \includegraphics[width=\textwidth]{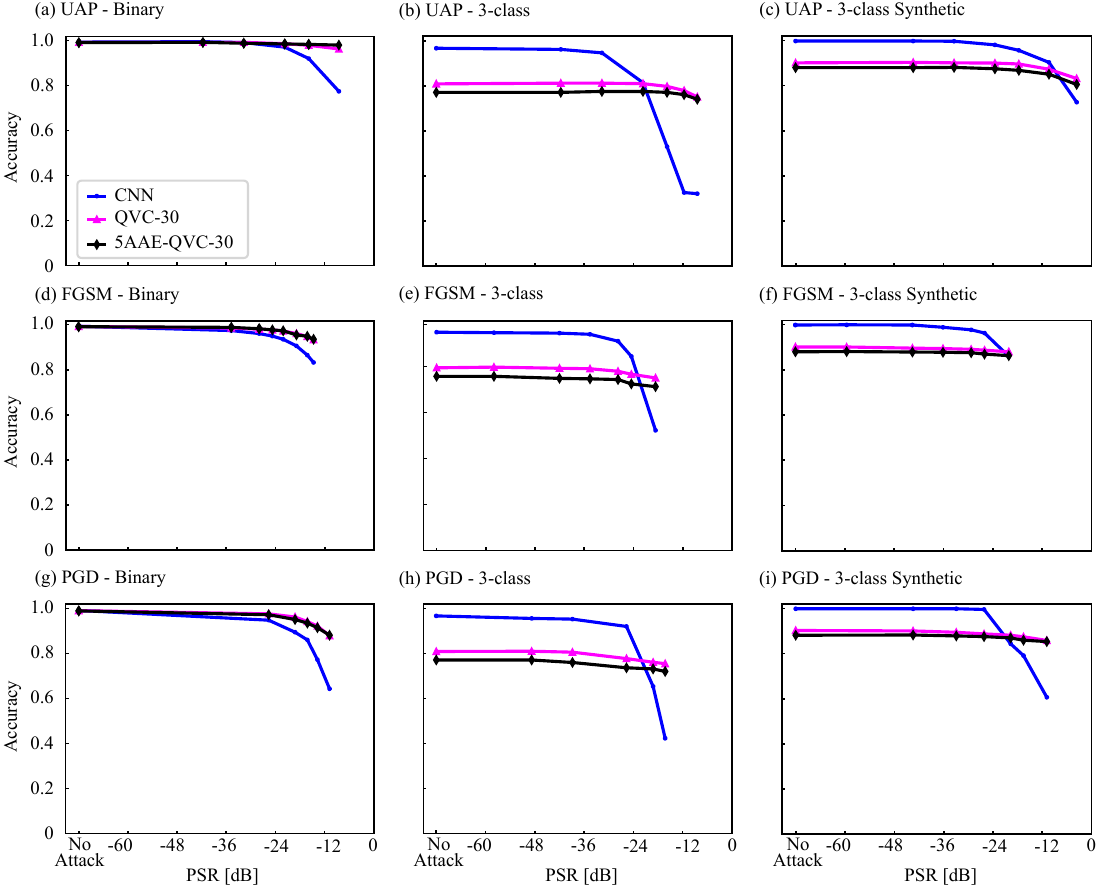}
    \caption{The performance achieved by QVC (purple), AAE-QVC (black), and CNN (blue) in the case of black-box UAP, FGSM, and PGD attacks on the GNU radio ML dataset and synthetic dataset as a function of attack strength defined as perturbation-to-signal ratio (PSR). Figures (a) to (c) represent the performance against black-box UAP attack for binary, 3-class RML dataset and synthetic dataset, respectively. Figures (d) to (f) represent the performance against black-box FGSM attacks for binary, 3-class RML dataset, and synthetic dataset, respectively. Figures (g) to (i) represent the performance under black-box PGD attack for binary RML dataset, 3-class RML dataset, and synthetic dataset, respectively. For each sub-figure, the left most dots represent the initial accuracy of the three models without any attacks. Across different attacks and datasets, we witness that attacks generated on quantum classifiers transfer well to classical CNN models and that the accuracy of the CNN models decreases sharply. These adversarial examples are capable of fooling the neural networks that they are not explicitly designed to attack. However, the converse is not true. Quantum classifiers largely resist the attacks generated on CNN models.}
\label{fig:bb-results}
\end{figure*}

\section{Datasets and Methods}
We compare the performance and robustness of QVC and CNN under the three types of attacks introduced in Section \ref{sec:adversarialattack} under both white-box and black-box scenarios. For the classical CNN model, we provide details of the model architecture and hyper-parameters used in this work in the Appendix \ref{hyper}. For the quantum classifier QVC, we provide details of the model in Section \ref{sec:qvc}. Both models are trained using the same GNU radio ML dataset RML2016.10a \citep{grcon} and our synthetic Fourier series waveform dataset.

\subsection{GNU Radio ML dataset}
The GNU radio ML (RML) dataset is publicly available at deepsig \footnote{\href{https://www.deepsig.ai/datasets}{https://www.deepsig.ai/datasets}}. It is a synthetic dataset generated with GNU Radio that consists of 10 modulations. 
GNU radio ML dataset is a variable-SNR dataset with moderate LO drift, light fading, and numerous labeled SNR increments for measuring performance across different signal and noise power scenarios \citep{grcon}. By offering a standardized dataset, RML dataset allows researchers to benchmark the performance of different machine learning algorithms and approaches in a consistent and fair manner. The variety and complexity of signals in the dataset also help in preparing machine learning models to handle real-world scenarios and challenges in RF environments. 
The dataset contains 200,000 signal samples, each belonging to a specific modulation class at a particular signal-to-noise ratio (SNR). The dataset is balanced. There are 10 modulation classes, and each contains the same number of signal samples for 20 different SNR levels from -20 dB to 18 dB with an interval step of 2 dB. Each sample holds 128 in-phase and 128 quadrature components as a 2$\times$128 matrix. 

Although not a limitation of our work, to reduce the computation burden of training large quantum classifiers, we randomly sample an equal number of signal data from two (8PSK vs. GFSK) and three (8PSK vs. GFSK vs. CPFSK) modulation classes at the SNR=16 dB to perform analysis on binary and 3-class classification respectively. Specifically, we sample 5000 data points per class for training and 500 per class for testing. The visualization of signal data for each modulation class can be found in Appendix \ref{app-visual}. We refer to this dataset as the RML dataset in the paper.  

\subsection{Synthetic Fourier Series dataset}
Fourier series can describe a periodic signal in terms of cosine and sine waves. In other words, it allows us to model any arbitrary periodic signal with a combination of sines and cosines \cite{zygmund02}. 
The motivation for creating this synthetic dataset is to verify the robustness of quantum classifiers to adversarial attacks in a general setting on a wide variety of signals that can be decomposed using the Fourier series. 
Specifically, as a proof of concept, this paper uses the Fourier series for three types of periodic signals: square waves, sawtooth waves, and triangle waves. For each type, we generate 5000 data points for training and 500 data points for testing. Each data point has added Gaussian noise and has varying frequency. The visualization of signal data for each class is shown in Appendix \ref{app-visual}. 

\subsection{Quantum Classifier}
\label{sec:qvc}
Quantum classifiers can only take quantum states as input. Thus, when classifying classical data using QVC, we first need to convert classical data into quantum states. There are two common ways of encoding classical data into quantum states: amplitude encoding and phase
encoding \citep{PhysRevA.102.032420}. In a $n$-qubit system, applying amplitude encoding can map an input vector $x \in \mathbb{R}^{p} $ directly into the amplitudes of the $2^n$ dimensional ket vector $|\varphi\rangle_\text{in}$ ($p \leq 2^n$). Amplitude encoding takes advantage of the exponentially large Hilbert space of the quantum computer; however, it requires exponentially deep encoding quantum circuits. In contrast, phase encoding is more efficient in terms of circuit depth, but it needs $p$ qubits to encode the input vector, seriously limiting the ability to encode complex datasets. 

For the RML dataset, the classical input data is of dimension $p=256$ in the form of vectors with 256 components. We chose amplitude encoding as our loading method as only 8 qubits are needed to encode all 256 components. We use two methods of amplitude encoding; the first is an exact encoding, where the required amplitudes are calculated directly from the classical data and loaded onto the qubits, and the second is an approximate amplitude encoding (AAE) method that relies on a parameterized circuit to approximate the correct amplitudes. 

After encoding the classical data, we need an optimizable quantum variational circuit (QVC) before measurements to determine the classifier's prediction. A QVC consists of a repeated layer of several parameterized single-qubit gates followed by two-qubit entanglement gates (see Fig. \ref{fig:architect}). The parameters of the single qubit gates are classically optimized to minimize the differences between the predicted class and the actual class, similar to classical neural networks. A more detailed discussion on hyperparameters and implementation details of our QVC is given in the Appendix \ref{hyper}.

\subsection{Approximate Amplitude Encoding}
\label{sec:aae}
Although amplitude encoding offers a highly efficient way of encoding data for QML, it is a well-known limitation that such encoding schemes require large circuit depths to implement. The number of gates required to load data into a quantum circuit scales as $2^{n}$ where $n$ is the number of qubits data is loaded onto \cite{shende2005synthesis}, meaning that unless we are working with very few qubits, the amplitude encoding process is likely to dominate runtimes in QML implementations. The average number of gates necessary for encoding our input data is almost 1000, while \cite{2023arXiv230909424W} found that roughly 2500 CNOT gates were needed to encode image data onto 9 qubits. Furthermore, unless we are working with very few qubits, the generic amplitude encoding process is prone to accumulating errors during physical implementation as a result of the high gate count, even before any computation is performed on the encoded data \cite{2021arXiv211206255Q}. 

One recent solution to this problem is to use an approximation of the true amplitude encoding process to encode data \cite{2023arXiv230909424W, nakaji22, zoufal19, 2023arXiv230913108J}. Instead of using exact amplitudes obtained with the usual amplitude encoding process to encode data, we use approximations of these exact amplitudes instead. The number of gates required to produce the approximate amplitudes is far smaller than the number required to produce the exact amplitudes (see, for example, \cite{2023arXiv230913108J,2023arXiv230909424W}). Although this encoding approach has inherent data loss, it offers a major advantage by dramatically reducing the circuit depth needed for encoding, thereby making it suitable for implementation and application in the NISQ era, where some degree of error or noise in the data is tolerable. Moreover, this will also significant reduce the amount of resources required for a fault-tolerant implementation \cite{West_2023}. 

\begin{figure*}
\centering
\begin{adjustbox}{width=1.02\textwidth}
\begin{quantikz}
\ket{0}&\gate{R_{X}(\theta_{000})}\gategroup[wires=7,steps=5,style={dashed, inner xsep=2pt}]{Layer 1}&\gate{R_{Z}(\theta_{001})}&\gate{R_{Y}(\theta_{002})}&\ctrl{1} & \phantomgate{0} & \phantomgate{0} & \ \ldots\ && \gate{R_{X}(\theta_{040})}\gategroup[wires=7,steps=5,style={dashed, inner xsep=2pt}]{Layer 5}&\gate{R_{Z}(\theta_{041})}&\gate{R_{Y}(\theta_{042})}&\ctrl{1} & \phantomgate{0} & \meter{} \\
\ket{0}& \gate{R_{X}(\theta_{100})}&\gate{R_{Z}(\theta_{101})}&\gate{R_{Y}(\theta_{102})}&\targ{} & \ctrl{1} & \phantomgate{0} & \ \ldots\ && \gate{R_{X}(\theta_{140})}&\gate{R_{Z}(\theta_{141})}&\gate{R_{Y}(\theta_{142})}&\targ{} & \ctrl{1} & \meter{} \\
\ket{0}& \gate{R_{X}(\theta_{200})}&\gate{R_{Z}(\theta_{201})}&\gate{R_{Y}(\theta_{202})} &\ctrl{1} & \targ{}& \phantomgate{0} & \ \ldots\ && \gate{R_{X}(\theta_{240})}&\gate{R_{Z}(\theta_{241})}&\gate{R_{Y}(\theta_{242})} &\ctrl{1} & \targ{}& \meter{} \\
& & & & & & & & & & & & & & & &\\
\vdots\ & & \vdots\ & & \vdots\ & & & \vdots\ & & & \vdots\ & & \vdots\ & & \vdots\ \\
\\
\ket{0}& \gate{R_{X}(\theta_{700})}&\gate{R_{Z}(\theta_{701})}&\gate{R_{Y}(\theta_{702})} & \targ{} & \phantomgate{0} & \phantomgate{0} & \ \ldots\ && \gate{R_{X}(\theta_{740})}&\gate{R_{Z}(\theta_{741})}&\gate{R_{Y}(\theta_{742})} & \targ{} &\phantomgate{0}& \meter{}
\end{quantikz}
\end{adjustbox}
\caption{The structure of the encoding QVC, $U(\bm{\theta})$, used to produce approximate amplitudes. It consists of 8 qubits and 5 variational layers of a $R_X$, $R_Z$ and $R_Y$ sequence, followed by a 2-qubit C-NOT gate. Each layer contains 3 trainable parameters $\theta_{ijk}$ per qubit/wire, where the subscripts $i$, $j$, and $k$ denote the qubit index, layer index, and gate index, respectively.}
\label{fig:AAE-QVC}
\end{figure*}
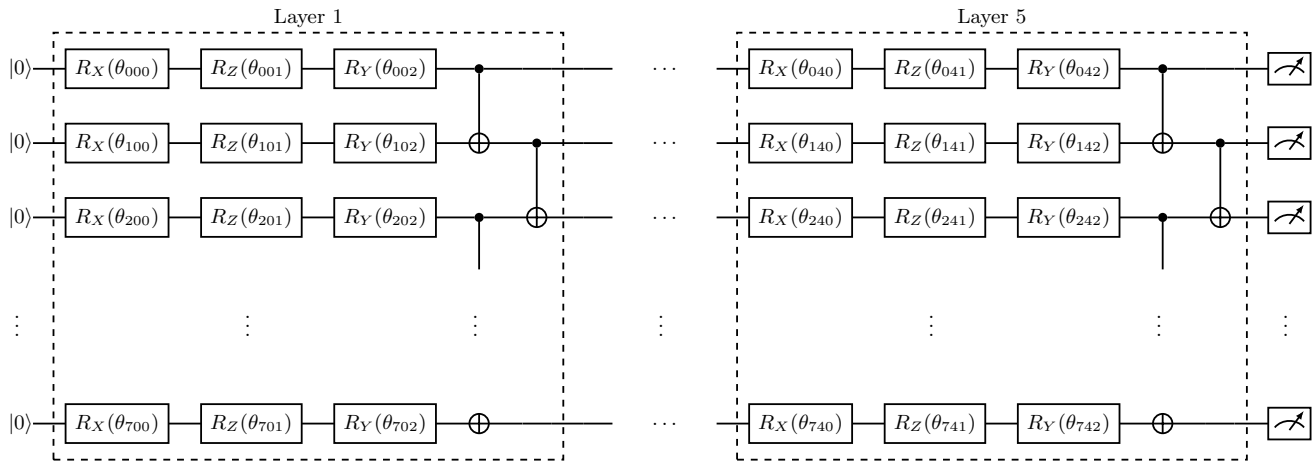

Similar to the method described in \cite{nakaji22}, the method of approximate amplitude encoding employed in this study involves the training of a shallow QVC to produce approximations of the amplitudes that would be obtained through the generic method of amplitude encoding (hereafter referred to as exact amplitude encoding). We use 8 qubits to encode the 256 components to the input signal data, where each qubit is initialized to the zero state. The variational component of the circuit is made up of $l$ layers where each layer consists of a sequence of 3 rotation gates per wire: a rotation about the X-axis ($R_X$), rotation about the Z-axis ($R_Z$) and a rotation about the Y-axis ($R_Y$), followed by a Controlled-NOT gate connecting two adjacent wires. The rotation angles of the $R_X$, $R_Z$, and $R_Y$ gates constitute the trainable parameters of the QVC (24 parameters per layer). For most of this study, we choose $l=5$ layers (equivalent to 360 trainable parameters), as this is sufficient to approximate the original amplitudes to a fidelity of approximately $0.89 - 0.90$. The structure of the circuit is depicted in Fig. \ref{fig:AAE-QVC}. 

The goal when training the encoding QVC was to train the parameters so that the output (``predictions") from the QVC matches as well as possible to exact amplitudes obtained from amplitude encoding (``labels"). Mathematically, we are producing approximate amplitudes through the application of the QVC to an initial zero state: 

\begin{equation}
U(\bm{\theta}) \ket{0}^{\otimes n} = \sum_{j=0}^{N-1} a_{j} \ket{j},
\label{AAE-eqn}
\end{equation}
where $U(\bm{\theta})$ represents the QVC to be trained, with trainable parameters given by the vector $\bm{\theta}$, $N$ is the number of available quantum states and is equal to $2^{n}$ where $n$ is the number of qubits that will encode the data, and \{$a_{j}$\} are the amplitudes, obtained from the approximate encoding, and associated with the quantum states given by $\ket{j} = \ket{j_{0}j_{1}...j_{N-1}}$ where $\ket{j_{k}}$ represents the state of the $k$-th qubit in the computational basis. Thus, the right-hand side of the equation above is the quantum state obtained after the approximate amplitude encoding process. 

We trained the parameters of $U(\bm{\theta})$ to best match the following criterion: 

\begin{equation}
|a_{j}|^{2} = |A_{j}|^{2}, \forall j \in [0, 1, ..., N-1], 
\label{opt}
\end{equation}
where \{$A_{j}$\} are the amplitudes obtained from exact amplitude encoding, and which the amplitudes \{$a_{j}$\} are approximating. However, we utilized an indirect method to optimize the above equation, involving training the inverse of the desired QVC, namely $U^{-1}(\overline{\bm{\theta}})$, to produce the initial input state $\ket{0}^{\otimes n}$, and using the desired output amplitudes as input to the inverse QVC. As such, the parameters $\overline{\bm{\theta}}$ of $U^{-1}(\overline{\bm{\theta}})$ were trained to satisfy the condition: 

\begin{equation}
|\bar{a}_{0}|^{2} = |\bar{A}_{0}|^{2} = 1, 
\label{rev-opt}
\end{equation}
where $\bar{a}_{0}$ and $\bar{A}_{0}$ are the approximated and exact amplitude of the state $\ket{0}^{\otimes n}$, which is equal to $1$ since the input state to $U(\bm{\theta})$ consists entirely of this zero state. Training $U^{-1}(\overline{\bm{\theta}})$ to satisfy (\ref{rev-opt}) is equivalent to training $U(\bm{\theta})$ to satisfy (\ref{opt}). The desired parameters $\bm{\theta}$ are obtained from the trained parameters $\overline{\bm{\theta}}$ by reversing the order of the components of the vector and flipping their signs. 

We used Pennylane's implementation of the Adam optimization algorithm, with a stepsize of 0.1 over 300 iterations. We also used values $\beta_{1} = 0.9$ and $\beta_{2} = 0.99$ for the hyper-parameters governing the updates of the first and second moments, and $\epsilon = 10^{-8}$ for the numerical stability offset, the details and explanations of which can be found in \cite{kingma14}. As mentioned above, we optimized based on the probabilities of measuring specific quantum states, with the view of implementing on quantum computers where state probabilities, not amplitudes, are observable and measurable. However, using probabilities does not allow global phases to be measured or approximated, meaning that the approximate amplitudes produced through this method include some unpredictable global phase given by $e^{i\phi}$. Thus, the approximate amplitudes may have both a real and imaginary component and may be flipped in sign relative to the original amplitude. For this study, we keep the global phase in the approximated amplitudes, as we find that removing the phase does not significantly impact the classification accuracies achieved by the signal-classifying QVC.

We use encoding fidelity as a measure of similarity between the approximate amplitudes and the exact amplitudes, defined as follows: 

\begin{equation}
    F = |\braket{\psi_{\rho}|\psi_{\sigma}}|^{2},
\end{equation}
where $F$ is the encoding fidelity, and $\psi_{\rho}$ and $\psi_{\sigma}$ represent the pure quantum states defined by approximate amplitudes and exact amplitudes, respectively. 

Although individual encodings for specific signals will vary with each run, we find that on average, the fidelities achieved with the approximate amplitudes vary only slightly per experiment. Using 5 encoding layers, we achieve a mean fidelity and standard deviation of $0.8957 \pm 0.0003$ from calculating approximate amplitudes for the 3-class RML dataset 10 times. As the variation is very small, we report the fidelities in our results section to 3 significant figures and do not report uncertainties in their value.

\section{Results and Discussion}
We present the results from (1) the application of QML to the well-known RML dataset (\cite{grcon}) and a synthetic Fourier Series dataset, and (2) the implementation of three different adversarial attacks (introduced in Section \ref{sec:adversarialattack}) on QVC and CNN models. In Section~\ref{sec:robustness}, we discuss the robustness of each of the three models (CNN, QVC and AAE-QVC) to adversarial attacks. 
We use the metric of perturbation-to-signal ratio (PSR) to measure the strength of the adversarial attacks, defined as the ratio of the perturbation power to the signal power: $\text{PSR}_{dB} = 10 \times \log (\text{P}/\text{S})$. In Section~\ref{sec:accuracy}, we compare the classification accuracies achieved by the three models and discuss the impact of the encoding circuit depth on the classification accuracy of the QVC model. We also report on the perceptibility of attacks generated by the QVC and CNN models in Section~\ref{sec:ablations}, and the effect of quantum noise on the accuracy and robustness of the models in Section~\ref{sec:noise}. 

\subsection{Robustness against adversarial attacks and attack transferability}
\label{sec:robustness}
In Fig. \ref{fig:wb-results}, we plot the performance achieved by the QVC and CNN models 
in the case of \emph{white-box} UAP, FGSM and PGD attacks on the RML dataset (first two columns) and our synthetic dataset (last column) as a function of PSR. We label the QVC models with the number of layers in the architecture, e.g., QVC30 consists of 30 layers. 
The performance of each model without adversarial attacks is presented in the figure by the left most dots. 
While QVC exhibits comparable accuracy to CNN in binary classification and slightly lower accuracy in 3-class classification, we find that, when subjected to the same PSR level and tested against three distinct types of adversarial attacks in a white-box scenario, QVC (highlighted in purple) generally demonstrates a slightly higher level of robustness compared to CNN. Nevertheless, as the PSR level increases, the accuracy of QVC still experiences a significant decline, thereby revealing its vulnerability to white-box adversarial attacks.

In Fig. \ref{fig:bb-results}, we show the accuracy achieved by QVC, AAE-QVC, and CNN in the case of \emph{black-box} UAP, FGSM, and PGD attacks on the same sets of datasets as a function of PSR. Similarly, the left most dots represent the initial accuracy of the algorithms without any attacks. Across different attack types and datasets, we witness that attacks generated on quantum classifiers exhibit strong transferability to classical CNN models, significantly reducing CNN accuracy. 
For instance, when evaluating the robustness against a black-box PGD attack on the 3-class RML dataset, as depicted in Figure (h), the CNN's accuracy plummets from 96.7\% to 43.4\%. These adversarial examples effectively deceive the neural networks they are not explicitly designed to attack. However, the converse scenario is not true. Quantum classifiers, namely QVC and AAE-QVC, show substantial resilience against attacks generated on convolutional neural networks. Their accuracy remains relatively stable even as the PSR level increases.

\subsection{Classification accuracy}
\label{sec:accuracy}

\begin{table}[]
    \centering
    \begin{tabular}{|c|c|c|c|}
    \hline
        Dataset & CNN & QVC-30 & 5AAE-QVC-30  \\
    \hline
        Binary RML & 0.995 & 0.993 & 0.992 \\
        3-class RML & 0.967 & 0.812 & 0.775 \\
        Synthetic & 0.998 & 0.902 & 0.881 \\
    \hline
    \end{tabular}
    \caption{Classification accuracy of CNN, QVC-30 and 5AAE-QVC-30 for binary RML, 3-class RML and the synthetic Fourier series datasets.}
    \label{tab:accuracy}
\end{table}

Table \ref{tab:accuracy} displays the classification accuracy attained by each of the CNN, QVC-30, and 5AAE-QVC-30 models. We tested the classification accuracy of each model on the binary RML dataset, the 3-class RML dataset, and the 3-class synthetic dataset. Across the different datasets, CNN achieves the highest classification accuracy, followed by QVC-30 and 5AAE-QVC-30. While the classification accuracy of QVC-30 matches the accuracy achieved by the CNN in the binary classification problem, we find that it achieves lower accuracies than CNN when classifying into 3 classes, using both the RML and synthetic datasets. This may result from insufficient complexity in the QVC model compared to the CNN model, which a multi-class problem would require. Increasing the number of layers or qubits in the QVC could improve the accuracy, though such changes will increase quantum resources on a quantum processor. In all datasets, the 5AAE-QVC-30 model accuracies are slightly lower than the QVC-30 accuracies due to the unavoidable data loss inherent in the amplitude approximation process. 

In Table~\ref{tab:AAE}, we report the encoding fidelities and classification accuracies for approximate encoding circuits with varying circuit depth. The dataset used for these results is the 3-class RML dataset, consisting of $16,500$ data samples. The number of gates per encoding layer is 31, so the total gate number for each experiment is the number of encoding layers multiplied by 31. We report baseline results for the exact encoding case for comparison, where the classification accuracy is $0.812$ when we use exact amplitudes as input to the QVC and where an average of $973$ gates (or roughly $121$ gates per qubit) is required to perform exact amplitude encoding across the dataset.  

We find that the encoding fidelity increases with a number of layers used in the approximate encoding circuit, which is unsurprising as having more trainable parameters tends to produce better approximations. 
Our results show that reasonably high accuracies can be achieved despite significantly reducing the number of gates used to encode the data. The success of the approximate encoding method is most apparent in the results obtained with 15 and 20 encoding layers. With only $465$ gates, less than half the number of gates necessary for exact encoding, we can achieve an accuracy of only $2.7\%$ below the peak classification accuracy obtained with exactly encoded amplitudes. This difference shrinks to below one percent when encoding with 20 layers, less than two-thirds of the number of gates required for exact encoding. The decision to use approximate encoding and the choice of layer depth, depends on the accuracy one is willing to accept for the problem at hand. For example, if a loss of accuracy of $3\%$ is acceptable, then we can use the much shallower encoding circuit to calculate the amplitudes. 

\begin{table}
    \centering
    \begin{tabular}{|c|c|c|c|}
    \hline
    Number of Approx. & Encoding & Classification & Total Number\\
    Encoding Layers & Fidelity & Accuracy & of Gates\\
    \hline
    3 & 0.731 & 0.746 & 93\\
    5 & 0.895 & 0.775 & 155\\
    10 & 0.944 & 0.779 & 310\\
    15 & 0.951 & 0.785 & 465\\
    20 & 0.988 & 0.803 & 620\\
    \hline
    Exact Encoding & 1.00 & 0.812 & 973\\
    \hline
    \end{tabular}
    \caption{Classification accuracy of 3-class RML dataset with QVC-30 using different approximate amplitude encoding (AAE) layers.}
    \label{tab:AAE}
\end{table}

\subsection{Data stealthiness}
\label{sec:ablations}
Attack stealthiness is an essential characteristic of adversarial attacks. A successful stealthy attack is considered as not perceptible and hard to detect. In this section, we propose a statistical analysis to compare the stealthiness of adversarial attacks generated by QVCs and classical CNNs. We adopt the assumption from \citep{10.1007/978-3-030-91431-8_57} to our scenario that an attack is imperceptible if we detect at least one modulation class that has a similar data distribution of radio signal data as that of the adversarially crafted radio signal data. For this purpose, we apply the Two-sample Kolmogorov-Smirnov goodness of fit test (KS-test). Basically, for an adversarial example $\Tilde{x}$, we apply the KS-test on every pair of $(\Tilde{x}, x_i)$ where $x_i$ is from the original dataset. KS-test will return the $p\_value$, which is compared with a significant level of $\alpha$. If $p\_value > \alpha$, we claim that the adversarial example is similar to the original data of at least one class. In practice, we set the significant level $\alpha$ to 0.05, which is the most commonly used value \citep{ijcai2019p647} and recommended as a standard level \citep{07b1346a-1f4b-37f4-ab7e-673fef1ea9e3}. A pseudo-code of the algorithm is presented in the Appendix \ref{app-ks}. In Table \ref{tab:ks}, we present the percentage of perceptible adversarial examples generated by QVCs and CNNs for 3-class RML datasets using FGSM, PGD, and UAP attacks under the white-box scenario. Under different PSR values for the FGSM and PGD attacks, adversarial examples generated by QVC-30 are considered more imperceptible than adversarial examples generated by CNN. For UAP attacks, adversarial examples generated by CNN are considered slightly more imperceptible than those generated by QVC. 
One potential explanation for the observed inconsistency in performance across various attack methods could be attributed to the distinct nature of UAP attack compared to the FGSM method and PGD method. UAP generates a single, universal perturbation noise applicable to all data points, whereas FGSM and PGD tailor a specific perturbation for each individual data point.
Additional perceptible perturbation analyses for the remaining scenarios are shown in the Appendix \ref{app-data-stealth}.

\begin{table}[h]
    \centering
    \caption{The percentage of out-of-distribution (detectable) adversarial examples generated by QVCs and CNNs for 3-class RML dataset under FGSM, PGD, and UAP attack. Under different PSR values for the FGSM and PGD attacks, adversarial examples generated by QVC-30 are considered generally more imperceptible (harder to detect) than adversarial examples generated by CNN. For UAP attack, adversarial examples generated by CNN are considered slightly more imperceptible than those generated by QVC.}
    \begin{tabular}{|c|c|c|c|c|c|}
    \hline
         PSR (FGSM attack) & -58 dB & -42 dB & -35 dB & -28 dB & -25 dB\\
    \hline
         QVC30 & \textbf{0.3\%} & \textbf{0.3\%} & \textbf{0.5\%} & \textbf{0.7\%} & \textbf{2.4\%} \\ 
    \hline
         CNN & 1.1\% & 2.1\% & 4.5\% & 6.7\% & 14.9\% \\
    \hline 
    
    \hhline{|=|=|=|=|=|=|}
        PSR (PGD attack) &  -48 dB & -38 dB & -25 dB & -18 dB & -15 dB\\
    \hline
        QVC30 & \textbf{0.3\%} & \textbf{0.3\%} & \textbf{0.6\%} & \textbf{1.7\%} & \textbf{2.4\%} \\
    \hline
        CNN & \textbf{0.3\%} & 2.1\% & 6.1\% & 18.5\% & 28.1\% \\
    \hline 

     \hhline{|=|=|=|=|=|=|}
        PSR (UAP attack) &  -42 dB & -32 dB & -22 dB & -16 dB & -12 dB\\
    \hline
        QVC30 & 0.3\% & 0.5\% & 2.7\% & 8.2\% & 13.1\% \\
    \hline
        CNN & \textbf{0.1\%} & \textbf{0.2\%} & \textbf{1.9\%} & \textbf{6.2\%} & \textbf{10.6\%} \\
    \hline
    \end{tabular}
    \label{tab:ks}
\end{table}

\begin{figure*}
    \centering
    \includegraphics[width=\textwidth]{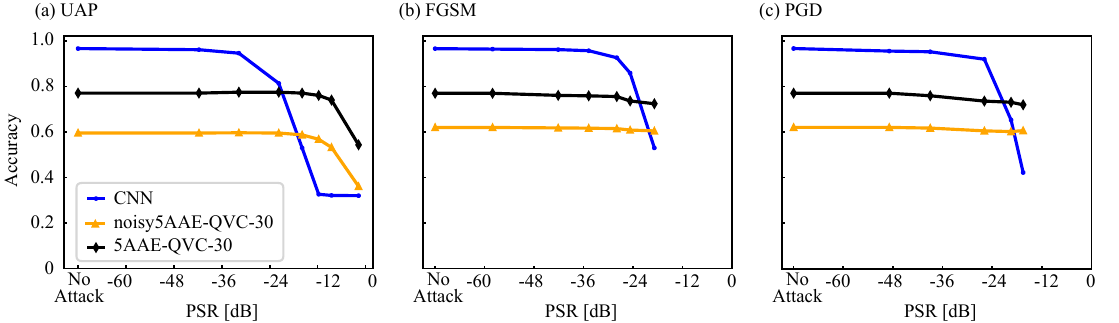}
    \caption{The performance achieved by AAE-QVC with depolarizing noise with a probability of 2\% (orange), AAE-QVC (black), and CNN (blue) in the case of black-box UAP, FGSM, and PGD attacks on the 3-class RML dataset. Figures (a) to (c) represent the performance against black-box UAP, FGSM, and PGD attacks, respectively. The left most dots represent the initial accuracy of the three models without any attacks.}
\label{fig:noise-results}
\end{figure*}

\subsection{Effects of Quantum Noise}
\label{sec:noise}
All experimental results of quantum classifiers presented are from noise-free quantum simulations. However, to run QML algorithms on a quantum machine, quantum noise is inevitable. To consider any effects of quantum noise on the robustness of QVCs and approximate amplitude encoding, we assess quantum noise channels on our AAE-QVC model and its robustness to adversarial attacks through noise simulations. The quantum noise channel we experiment with is a depolarizing error. Depolarizing error is modelled by the following Kraus matrices \cite{nielsen00}:
\begin{align*}
    K_0 = \sqrt{1-p}\left[ \begin{array}{cc}
1 & 0  \\ 0 & 1 \end{array} \right] \\
    K_1 = \sqrt{p/3}\left[ \begin{array}{cc}
0 & 1  \\ 1 & 0 \end{array} \right] \\
    K_2 = \sqrt{p/3}\left[ \begin{array}{cc}
0 & -i  \\ i & 0 \end{array} \right] \\
    K_3 = \sqrt{p/3}\left[ \begin{array}{cc}
1 & 0  \\ 0 & -1 \end{array} \right] 
\end{align*}
where $p \in [0, 1]$ is the depolarization probability and is equally divided in applying all Pauli operations. We use a depolarization probability of 0.02 on every qubit and AAE and QVC layer to simulate this noise channel. We train the QVC with approximate amplitude encoding and this noise channel on the same 3-class RML radio dataset. The performance is presented in Fig. \ref{fig:noise-results}. Based on the results, it is evident that introducing depolarization noise with a probability of 0.02 leads to a significant drop in classification accuracy; however, it does not affect the QVC model's robustness towards the black-box adversarial attacks generated from the classical CNN model.

\section{Conclusions}
In this work, we performed a systematic and extensive assessment of the robustness of QVCs in the presence of various adversarial attacks and in comparison of the classical CNN model on the RML dataset and synthetic Fourier series waveform dataset. In addition, we propose the novel application of approximate amplitude encoding technique on radio signal data.  
Based on our experimental results, we witness several important insights of the QAML in applying radio signal classification. To the best of our knowledge, this is the first work to use QML and QAML for radio signal classification. Previous research mainly focused on the field of computer vision for image classification, hence, our work in signal data classification helps to bridge knowledge gaps in QML and QAML by answering key questions such as how well QML algorithms fare against strong classical adversarial attacks for radio signal datasets, will the robustness and attack transferability of QAML in signal data classification behave similarly as in image classification, and whether the perturbated radio signal data generated by attacking quantum models are more imperceptible than data generated by attacking classical classifiers. Our research will lay the foundation for the future development of QAML, and based on the insights we provide, some future work directions are worth looking at: 
\begin{itemize}
    \item In this work, we have benchmarked QVCs under various classical adversarial attacks and discovered the robustness of QVCs in the black-box scenario. However, in the white-box attacks, QVCs are still vulnerable to these attacks. An important future research direction of QAML is to develop defense strategies. West et al. \cite{PhysRevResearch.5.023186} found that adversarial training of QVCs produces relatively small improvements in accuracy compared to those in classical AML; hence, developing effective defense strategies for QAML is of great importance and secure robustness of quantum networks against both white-box and black-box attacks.
    \item While our work has focused on studying the robustness of QVCs against adversarial attacks, one important work direction would be to improve further the learning accuracy of quantum ML networks, especially for multi-class classification. 
    \item Another possible research direction is to look at a hybrid setting where we combine quantum and classical ML techniques to study whether we can reach high learning accuracy as in classical models for complex data and strong robustness against adversarial attacks in quantum models using just a few number of qubits and gates. 
\end{itemize}

In summary, vulnerability to intentionally crafted adversarial attacks has recently raised major security concerns for classical and quantum ML algorithms. This paper addresses a significant gap in the QAML literature by thoroughly studying the robustness and attack transferability of adversarial examples between classical and quantum neural networks in the context of radio frequency datasets. By discovering the failure of the classical adversarial examples to transfer to the QVCs in signal classification, we witness that the quantum ML models are generally impervious to the adversarial attacks that target the precise features learned by the classical networks. Such a scenario will offer a new form of advantage in QML. Moreover, by studying the data stealthiness of the adversarial examples, we discovered that, for attackers, adversarial attacks generated on quantum models using FGSM and PGD methods are generally more imperceptible. Quantum ML classifiers may not necessarily be more accurate than their classical counterparts, but they provide superior robustness to adversarial attacks and can generate more imperceptible attacks. 

\section*{Acknowledgements}
The authors acknowledge the support of the CSIRO Future Science Platforms program. This research/project was undertaken with the assistance of resources and services from the CSIRO's High Performance Computing cluster (Petrichor), as well as the National Computational Infrastructure (NCI), which is supported by the Australian Government. 

\bibliographystyle{unsrt}
\bibliography{qvc}

\newpage
\clearpage

\onecolumngrid
\appendix

\section{Hyperparameters and implementation details}
\label{hyper}

Table \ref{tab:hyper-parameter-QVC} and Table \ref{tab:hyper-parameter-cnn} give a list of hyperparameters used in the experiments. Hyperparameters are made the same across different attack scenarios to ensure fairness and consistency in comparisons. Our convolutional neural network begins with one layer of 1$\times$3 filter with 32 feature maps and ReLU activation function. The convolutional layer is followed by two fully connected layers, which utilize the ReLU activation function. The dropout rate is set to 0.2.  
Table \ref{tab:num-parameter} lists the number of parameters required for the different models considered. Quantum variational circuits require drastically fewer trainable parameters than their classical counterpart. 

\begin{table}[ht]
    \centering
    \caption{Hyperparameter values for QVC}
    \vskip 0.1in
    \begin{tabular}{l|c }
    \hline
         Hyperparameters & QVC \\
    \hline
         optimizer & Adam \\
         learning rate & $5 \cdot 10^\text{$-$3}$ \\
         number of QVC layers & 30 \\
         batch size & 256 \\
    \hline
    \end{tabular}
    \label{tab:hyper-parameter-QVC}
\end{table}

\begin{table}[ht]
    \centering
    \caption{Hyperparameter values for CNN}
    \vskip 0.1in
    \begin{tabular}{l|c }
    \hline
        Hyperparameters & CNN \\
    \hline
        optimizer & Adam \\
        learning rate & $1 \cdot 10^\text{$-$3}$ \\
         
        batch size & 256 \\
        Nonlinearity & ReLU \\
        Padding layer & (0, 2) \\
        Conv layer & (32, 1, 3) \\
        Dropout & 0.2 \\
        Flatten & - \\
        FC & (16) \\
        FC & (2) or (3)\\
    \hline 
    \end{tabular}
    \label{tab:hyper-parameter-cnn}
\end{table}

\begin{table}[ht]
    \centering
    \caption{The number of parameters required for QVCs and CNNs.}
    \vskip 0.1in
    \begin{tabular}{|c|c|c||c|c|}
    \hline 
         & \multicolumn{2}{c|}{Binary} & \multicolumn{2}{c|}{3-class}  \\
    \hline
         Network & Qubits & Parameters/Gates  & Qubits & Parameters/Gates \\
    \hline
         CNN & - & 133,298  & - & 133,315  \\
    \hline
         QVC-30 & 8 & 1,933 & 8 & 1,933  \\
    \hline 
         5AAE-QVC-30 & 8 & 1,115 & 8 & 1,115 \\
    \hline
    \end{tabular}
    \label{tab:num-parameter}
\end{table}

\clearpage
\section{Two-sample Kolmogorov-Smirnov goodness of fit test}
\label{app-ks}

In Algorithm \ref{alg:kstest}, we provide the pseudo-code for the Two-sample Kolmogorov-Smirnov goodness of fit test (KS-test) \citep{10.1007/978-3-030-91431-8_57} applied in Section \ref{sec:ablations}. We set the significant level $\alpha$ to 0.05, which is the most commonly used value \citep{ijcai2019p647} and recommended as a standard level \citep{07b1346a-1f4b-37f4-ab7e-673fef1ea9e3}.

\begin{algorithm}[H]
\caption{Two sample KS-test}
\label{alg:kstest}
\hspace*{\algorithmicindent} \textbf{Input:} $X_\text{adversarial}$, $X_\text{original}$, significant level $\alpha$
\begin{algorithmic}[1]
\For {$S_i$ in $X_\text{adversarial}$}
    \For {$S_j$ in $X_\text{original}$}
        \State KS\_test = Ks($S_i$, $S_j$)
        \If {$\text{Ks}_{p\_value} > \alpha$}
            \State $S_i$ is considered imperceptible
        \EndIf
    \EndFor
\EndFor
\end{algorithmic}
\end{algorithm}

\clearpage
\section{Visualization of datasets}
\label{app-visual}

\begin{figure}[h]
    \centering
    \includegraphics[width=\textwidth]{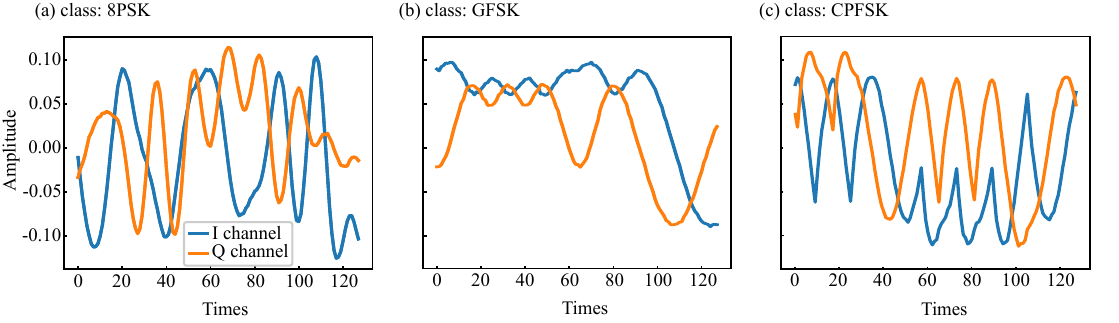}
    \caption{Visualization of GNU Radio ML dataset for three different modulation classes at the SNR=16 dB. }
    \label{fig:gnu}
\end{figure}

\begin{figure}[h]
    \centering
    \includegraphics[width=\textwidth]{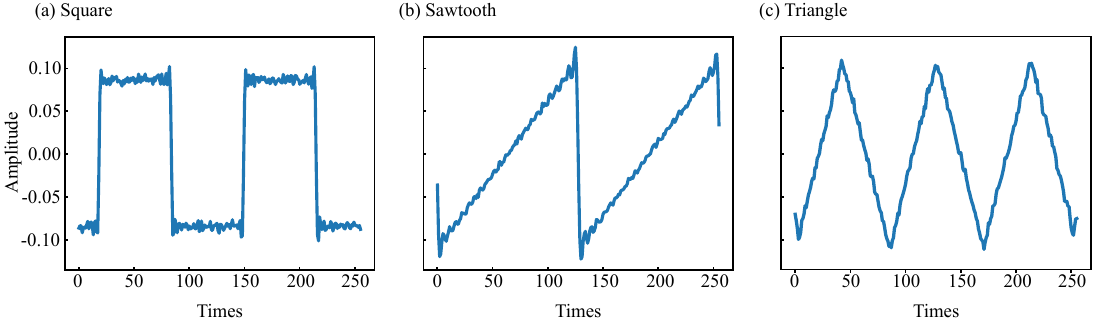}
    \caption{Visualization of Synthetic Fourier Series dataset. Three different types of periodic signals, square waves, sawtooth waves, and triangle waves, are generated in the dataset. Each data point has added Gaussian noise and has varying frequency. Figures (a) to (c) show the visualization of the three different periodic signals, respectively.}
    \label{fig:syn}
\end{figure}

\clearpage

\section{Visualization of Adversarial Examples}

\begin{figure}[h]
    \centering
    \includegraphics[width=\textwidth]{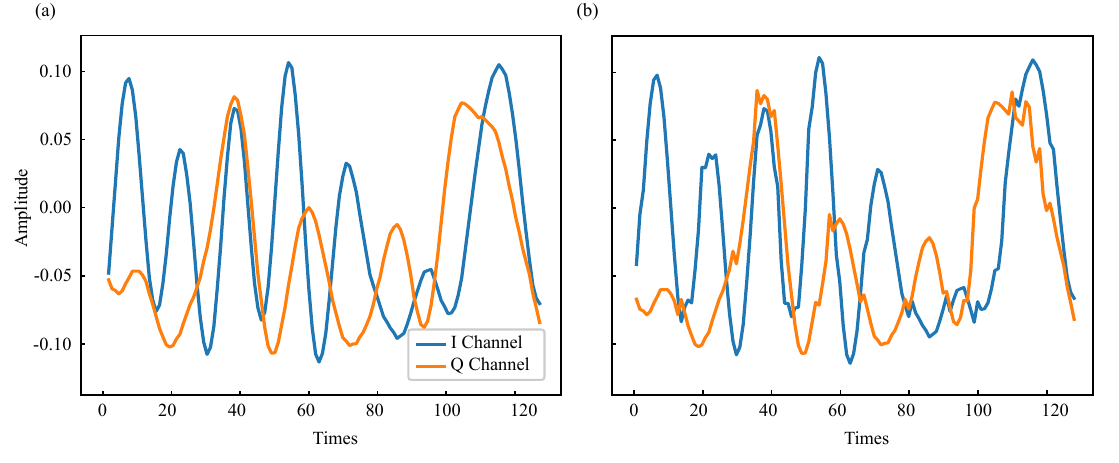}
    \caption{Visualization of an FGSM attack with PSR = -28[dB] on signal data with modulation class equals 8PSK. The original signal data is shown on the left. The perturbated signal data is shown on the right.}
    \label{fig:adv-examples}
\end{figure}

\section{Data Stealthiness}
\label{app-data-stealth}

\begin{table}[h]
    \centering
    \caption{The percentage of out-of-distribution (detectable) adversarial examples generated by QVCs and CNNs for binary RML dataset under FGSM and PGD attack. Under different PSR values for the FGSM and PGD attacks, adversarial examples generated by QVC-30 are considered generally more imperceptible (harder to detect) than adversarial examples generated by CNN. }
    \begin{tabular}{|c|c|c|c|c|c|}
    \hline
         PSR (FGSM attack) & -58 dB & -42 dB & -35 dB & -28 dB & -25 dB\\
    \hline
         QVC30 & 0.1\% & 0.6\% & 1.1\% & 1.8\% & 2.8\% \\ 
    \hline
         CNN & 2.3\% & 4.7\% & 6.8\% & 7.2\% & 9.3\% \\
    \hline 
    
    \hhline{|=|=|=|=|=|=|}
        PSR (PGD attack) &  -48 dB & -38 dB & -25 dB & -18 dB & -15 dB\\
    \hline
        QVC30 & 0.6\% & 2.9\% & 5.8\% & 8.1\% & 13.8\% \\
    \hline
        CNN & 4.4\% & 6.3\% & 7.7\% & 10.4\% & 16.6\% \\
    \hline 

    \hhline{|=|=|=|=|=|=|}
        PSR (UAP attack) &  -42 dB & -32 dB & -22 dB & -16 dB & -12 dB\\
    \hline
        QVC30 & 0.1\% & 0.4\% & 3.2\% & 7.3\% & 10.1\% \\
    \hline
        CNN & 0.1\% & 0.2\% & 1.9\% & 6.2\% & 8.6\% \\
    \hline
    \end{tabular}
    \label{tab:ks2}
\end{table}

\clearpage

\end{document}